# High-Performance HZO/InAlN/GaN MIS-HEMT with $f_T$/$f_{max}$ of 155/250 GHz

Peng Cui, Hang Chen, John Q. Xiao, and Yuping Zeng

*Abstract*—Scaling of GaN high-electron-mobility transistors (HEMTs) usually increases gate leakage current and deteriorates breakdown characteristic, limiting the maximum drain current and output power density. These bottlenecks can be circumvented by inserting a dielectric material under the gate of HEMTs. Doped $HfO_2$ is an excellent dielectric material but unexplored so far as the gate material of HEMTs for high-speed device application. Here we demonstrate that Zr-doped $HfO_2$ (HZO)-gated InAlN/GaN metal-insulator-semiconductor (MIS) HEMTs exhibit remarkable properties. The device with a gate length ($L_g$) of 50 nm exhibits maximum drain current ($I_{d,max}$) of 2.15 A/mm, a transconductance ($g_m$) peak of 476 mS/mm, an on/off current ratio ($I_{on}/I_{off}$) of $9.3 \times 10^7$, a low drain-induced barrier lowing (DIBL) of 45 mV/V. RF characterizations reveal a current gain cutoff frequency ($f_T$) of 155 GHz and a maximum oscillation frequency ($f_{max}$) of 250 GHz, resulting in a $(f_T \times f_{max})^{1/2}$ of 197 GHz. The breakdown voltages (BV) of 35 V and 72 V is achieved on the $L_g$ = 50 nm devices with source-drain distance ($L_{sd}$) of 0.6 and 2 μm ($f_T$ of 155 and 110 GHz), resulting in high Johnson's figure-of-merit (JFOM = $f_T \times$ BV) of 5.4 and 7.9 THz·V, respectively. These properties, particularly the high $f_T$/$f_{max}$ and JFOM are highly desirable for the millimeter-wave power applications, demonstrating the great technological potential of HZO/InAlN/GaN MIS-HEMTs.

## I. INTRODUCTION

GaN-based high-electron-mobility transistors (HEMTs) indicate great potential for RF and millimeter-wave power applications [1-5]. To date, excellent current gain cutoff frequency ($f_T$) and maximum oscillation frequency ($f_{max}$) have been demonstrated on GaN HEMTs with device scaling [6-9]. However, device scaling usually causes high gate leakage current and deteriorates breakdown voltage (BV), thus limiting the maximum drain current and output power density. The gate dielectric can suppress the leakage current and enhance the breakdown characteristic. Therefore, the introduction of gate dielectric on GaN metal-insulator-semiconductor HEMTs (MIS-HEMTs) could lead to further improvement of the device performance for high-speed and high-power applications.

For high-speed device application, different dielectric materials ($Al_2O_3$ [10-19], $HfO_2$ [20-22], SiN [23-27], $SiO_2$ [28], $TiO_2$ [29], MgCaO [30], ZnO [31], *et al.*) have been investigated as the gate dielectric in GaN MIS-HEMTs. The relevant device performance, such as maximum drain current ($I_{d,max}$) of 2.4 A/mm [32], on/off current ratio ($I_{on}/I_{off}$) of $5 \times 10^8$ [30], transconductance ($g_m$) of 653 mS/mm [11], $f_T$/$f_{max}$ of 190/300 GHz [20], and Johnson's figure-of-merit (JFOM) of 10.8 THz·V [23] has been demonstrated. Although there is still a gap between MIS-HEMTs and HEMTs, the advantages of MIS-HEMTs have been shown.

In this study, we report the first demonstration of the $Hf_{0.5}Zr_{0.5}O_2$ (HZO) as the gate dielectric for GaN-based high-speed MIS-HEMTs. Key device performance parameters, including high $I_{d,max}$, low drain-induced barrier lowing (DIBL), high $f_T$/$f_{max}$ and JFOM, are simultaneously achieved on the HZO/InAlN/GaN MIS-HEMTs, suggesting their great potential for high-speed applications.

## II. EXPERIMENT

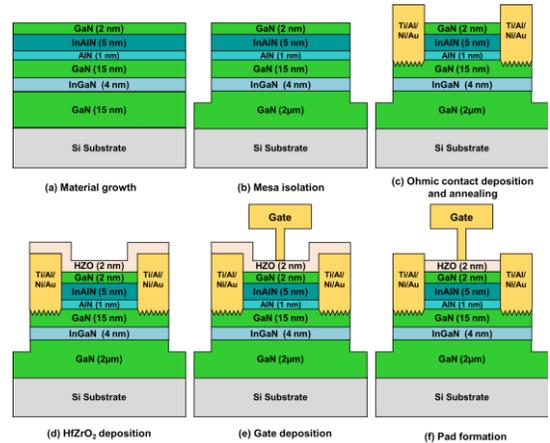

**Fig. 1**. Key process flow for the fabricated HZO/InAlN/GaN HEMT: **(a)** material growth, **(b)** mesa isolation, **(c)** ohmic contact deposition and annealing, **(d)** HZO gate dielectric deposition, **(e)** gate deposition, and **(f)** pad formation.

**Fig. 1** shows the key process flow of the fabricated HZO/InAlN/GaN MIS-HEMT. The growth of epitaxial structures is performed with metalorganic chemical vapor deposition (MOCVD) on a 4-inch high-resistance Si substrate. The epitaxial layer consists of a 2-μm undoped GaN buffer layer, a 4-nm $In_{0.12}Ga_{0.88}N$ back-barrier layer, a 15-nm GaN channel layer, a 1-nm AlN interlayer, a 5-nm lattice-matched $In_{0.17}Al_{0.83}N$ barrier layer, and a 2-nm GaN cap layer. Device mesa isolation was carried out with $Cl_2$-based inductively coupled plasma (ICP) etching with an etch depth of ~300 nm. Then ohmic contact was formed with Ti/Al/Ni/Au deposition and annealing at 850°C for 40 s. Then HZO was deposited as the gate dielectric and passivation layer by using plasma-enhanced atomic layer deposition (PEALD) at 150°C.

This work was supported in part by the NASA International Space Station under Grant 80NSSC20M0142, and in part by Air Force Office of Scientific Research under Grant FA9550-19-1-0297 and Grant FA9550-21-1-0076.

P. Cui and Y. Zeng are with the Department of Electrical and Computer Engineering, University of Delaware, Newark, DE, 19716, USA (e-mail: yzeng@udel.edu).

H. Chen and J. Q. Xiao are with the Department of Physics and Astronomy, University of Delaware, Newark, DE 19716, USA.

Tetrakis(dimethylamino)hafnium (TDMAH), Bis(methyl-η5−cyclopentadienyl)methoxymethylzirconium (ZRCMMM), and oxygen are used as Hf, Zr, and O source, respectively. The film was grown in a Hf: Zr ratio of 1:1 by alternating cycles of TDMAH, $O_2$, ZRCMMM, $O_2$. The alternating cycles were repeated 30 times for 2-nm HZO growth. These steps were followed by T-shaped gate fabrication with electron beam lithography and Ni/Au deposition. Finally, HZO on the pad was removed by dipping the samples in HF solution (HF: $H_2O$ = 1: 9) for 30s. Devices with source-drain distance ($L_{sd}$) of 2 ~ 0.6 μm, gate length ($L_g$) of 50 ~ 150 nm, and gate width ($W_g$) of 20 × 2 μm were fabricated.

## III. RESULTS AND DISCUSSION

Hall measurements were carried out on the InAlN/GaN heterostructure before and after HZO deposition. Before HZO deposition, the electron density ($n_{2d}$) of $1.71 \times 10^{13}$ cm$^{-2}$ and electron mobility ($\mu_{2d}$) of 1663 cm$^2$/V·s are obtained, an indication of a good InAlN/GaN heterostructure. After HZO deposition, $n_{2d}$ of $2.24 \times 10^{13}$ cm$^{-2}$ and $\mu_{2d}$ of 1613 cm$^2$/V·s are determined. The increased $n_{2d}$ presents a good passivation effect on the material surface [33, 34], and the negligible change in $\mu_{2d}$ means that the electron mobility is not degraded with the dielectric deposition. **Fig. 2(a)** depicts the diode curves, which show a ~ 4 order decrease of gate leakage current at a gate-source voltage ($V_{gs}$) of -10 V with HZO deposition. **Fig. 2(b)** presents the extracted interface trap density ($D_{it}$) of the HZO/InAlN/GaN diode by using the conventional conductance method [35, 36]. The inset of **Fig. 2(b)** plots the measured and fitted $G_P/\omega$ versus $\omega$ ($G_P$ is the measured conductance and $\omega$ is the radial frequency [35, 36]). The device shows a low $D_{it}$ of 1~3 ×10$^{12}$ eV$^{-1}$·cm$^{-2}$. The Fat-FET (HZO/InAlN/GaN with $L_g$ of 96 μm and $L_{sd}$ of 100 μm) is fabricated for low-field mobility extraction [37, 38]. **Fig. 2(c)** shows the measured capacitance and $n_{2d}$ at a 1MHz frequency of the Fat-FET. **Fig. 2(d)** exhibits low-field mobility (at $V_{ds}$ = 0.1 V) versus $n_{2d}$ of the Fat-FET, indicating a peak mobility of 1627 cm$^2$/V·s. All these properties unequivocally demonstrate that HZO a strong candidate as a gate dielectric for GaN MIS-HEMTs.

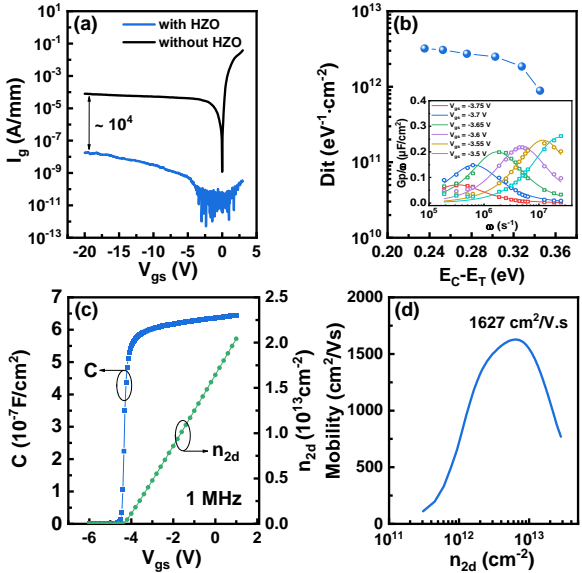

**Fig. 2(a)** $I_g$-$V_{gs}$ curves for the InAlN/GaN and HZO/InAlN/GaN diode. **(b)** Extracted $D_{it}$ as a function of $E_c$-$E_T$. Insert: Measured and fitted $G_P/\omega$ versus $\omega$. **(c)** Capacitance (C, left) and electron density ($n_{2d}$, right) versus $V_{gs}$. **(d)** Extracted low-field mobility versus $n_{2d}$ with a peak value of 1627 cm$^2$/V·s on an HZO/InAlN/GaN MIS-HEMT with $L_g$ = 96 μm and $L_{sd}$ = 100 μm.

**Fig. 3(a)** shows the typical output characteristic of the HZO/InAlN/GaN MIS-HEMT ($L_g$ = 50 nm, and $L_{sd}$ = 0.6 μm), depicting an on-resistance ($R_{on}$) of 1.41 Ω·mm. The transfer characteristic and transconductance ($g_m$) of the same device are shown in **Fig. 3(b)**. A maximum drain current ($I_{d,max}$) of 2.15 A/mm and a $g_m$ peak of 476 mS/mm are observed. The transfer and gate current characteristics in semi-log scale at $V_{ds}$ = 10 V and 1 V (**Fig. 3(c)**) exhibit a low leakage current, an on/off current ratio ($I_{on}/I_{off}$) of $9.3 \times 10^7$, a subthreshold swing (SS) of 130 mV/dec, and a drain-induced barrier lowing (DIBL) of 45 mV/V at $I_d$ = 10 mA/mm. With gate dielectric deposition, the low DIBL confirms insignificant short-channel effects (SCEs) with an aspect ratio of ~5 [1, 39]. **Fig. 3(d)** shows the off-state three-terminal breakdown characteristics of the $L_g$ = 50 nm HZO/InAlN/GaN MIS-HEMTs. The breakdown voltage (BV) of 72 V and 35 V is demonstrated on the device with $L_{sd}$ of 2 and 0.6 μm, respectively.

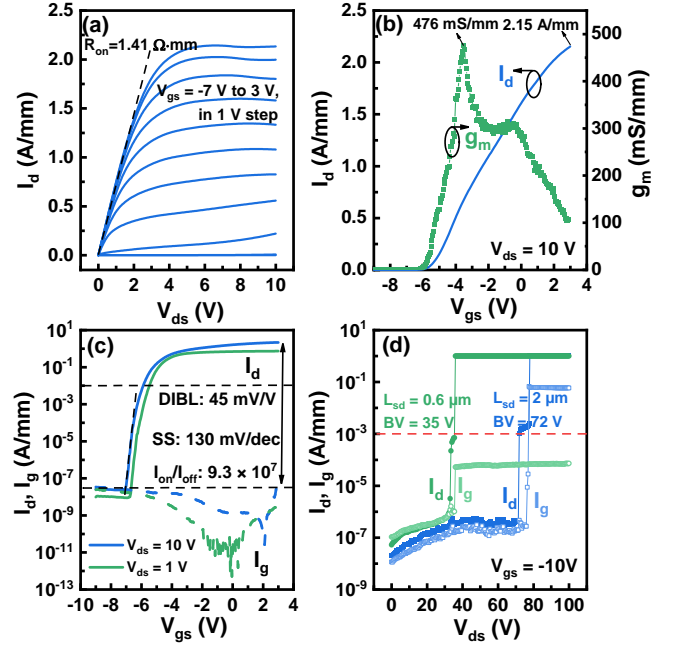

**Fig. 3** (a) Output characteristic, (b) transfer characteristic in linear scale (left), transconductance ($g_m$, right), and (c) transfer and gate current characteristics in semi-log scale at $V_{ds}$ of 10 V and 1V of the $L_g$ = 50 nm HZO/InAlN/GaN MIS-HEMT. (d) Off-state three-terminal breakdown characteristics for the $L_g$ = 50 nm HZO/InAlN/GaN MIS-HEMTs with $L_{sd}$ of 2 and 0.6 μm, respectively.

The microwave characteristics of the HZO/InAlN/GaN MIS-HEMTs are characterized from 1 to 65 GHz using an Anritsu MS4647B vector network analyzer. By using the de-embedded S-parameters, the high-frequency gains of the devices are extracted. **Fig. 4(a)** plots the measured short-circuit current gain ($|h_{21}|^2$), Mason's unilateral gain (U), maximum-stable-gain (MSG), and stability-factor (k) of the MIS-HEMT with $L_g$ of 50 nm and $L_{sd}$ of 0.6 μm at $V_{ds}$ = 10 V and $V_{gs}$ = -3.8 V. $f_T/f_{max}$ of 155/250 GHz is obtained by extrapolation of $|h_{21}|^2$ and U with a -20 dB/dec slope, resulting in $f_T \times L_g$ of 7.75 GHz·μm and $(f_T \times f_{max})^{1/2}$ of 197 GHz. $f_T/f_{max}$ versus $I_d$ is also measured and plotted in **Fig. 4(b)**. The classical 16-element equivalent-circuit model is used for the device, as shown in **Fig. 4(c)** [40, 41]. Based on the model, the device extrinsic and intrinsic parameters are extracted in **Fig. 4(d)** and the simulated $f_{T,model}$/

$f_{max,model}$ of 156/249 GHz are consistent with the measured results [40-42]. **Fig. 4 (e)** and **(f)** show the measured $f_T/f_{max}$ as a function of $L_g$ and $L_{sd}$, respectively. $f_T$ for the $L_g$ = 50 nm devices with $L_{sd}$ of 2 and 0.6 µm is 110 and 155 GHz (BV of 72 V and 35 V), resulting in the high Johnson's figure-of-merit (JFOM = $f_T \times$ BV) of 7.9 and 5.4 THz·V, respectively. **Fig. 5(a)** and **(b)** show the $f_{max}$ and B$V$ versus $f_T$ benchmark for the presented devices against state-of-the-art GaN MIS-HEMTs on SiC, Sapphire, Si, and GaN substrates [10-32]. The HZO/InAlN/GaN MIS-HEMTs on Si in this work exhibit excellent device performance on high $f_T/f_{max}$ and BV simultaneously, indicating the outstanding potential for high-speed and high power applications.

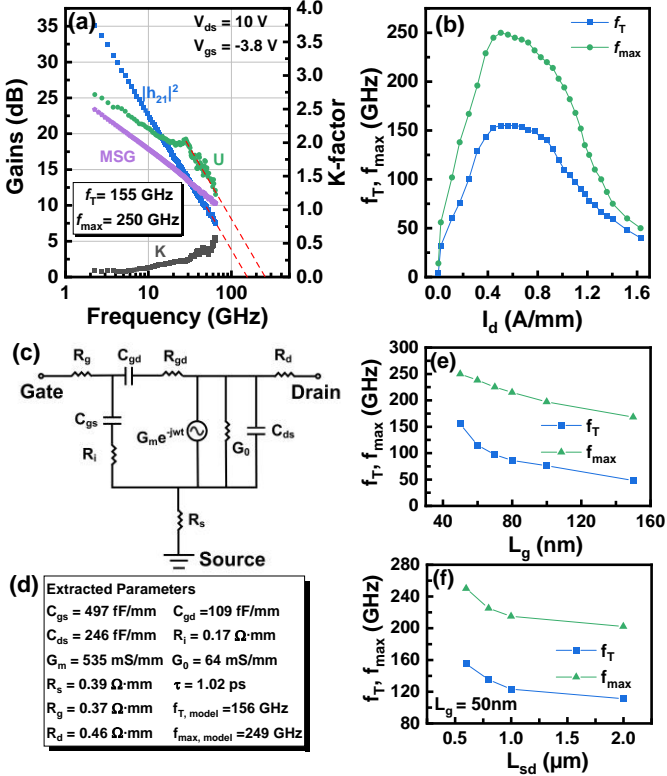

**Fig. 4 (a)** High-frequency gains ($|h_{21}|^2$, U and MSG), stability factor (k), **(b)** $f_T/f_{max}$ versus $I_d$, **(c)** small-signal equivalent-circuit model; and **(d)** the extracted intrinsic parameters of the $L_g$ = 50 nm HZO/InAlN/GaN MIS-HEMT. **(e)** and **(f)** $f_T/f_{max}$ of HZO/InAlN/GaN MIS-HEMT as a function of $L_g$ and $L_{sd}$, respectively.

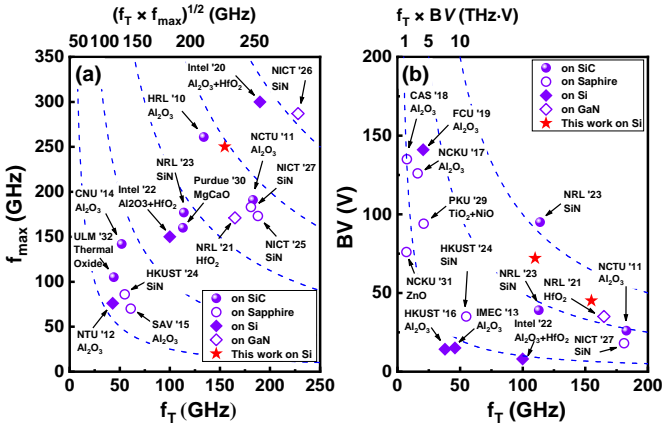

**Fig. 5 (a)** $f_{max}$ and **(b)** BV versus $f_T$ benchmark for the presented devices (HZO/InAlN/GaN MIS-HEMTs on Si) against state-of-the-art GaN MIS-HEMTs on SiC, Sapphire, Si, and GaN substrates.

## IV. CONCLUSION

In summary, by using HZO as the gate dielectric, the $L_g$ = 50 nm InAlN/GaN MIS-HEMT presents a high performance with $I_{ds,max}$ of 2.15 A/mm, $g_m$ peak of 476 mS/mm, $I_{on}/I_{off}$ of 9.3 × $10^7$, DIBL of 45 mV/V, $f_T/f_{max}$ of 155/250 GHz, and $(f_T \times f_{max})^{1/2}$ of 197 GHz. The devices with $L_{sd}$ of of 2 and 0.6 µm present high JFOM of 7.9 and 5.4 THz·V, respectively. The simultaneously achieved excellent cutoff frequencies and breakdown characteristics indicate the great potential of the HZO/InAlN/GaN MIS-HEMTs for RF and millimeter wave power applications.